\documentstyle[twocolumn,aps,epsf]{revtex}

\begin{document}

\title{Effects of Coulomb interaction and tunneling on electron
transport\\ in coupled one-dimensional systems:  from ballistic
to diffusive regime.}

\author{ O. E. Raichev\cite{oleg} and P. Vasilopoulos\cite{takis}}
\voffset 1cm
\address{\cite{oleg}Inst. of Semiconductor Physics, NAS Ukraine,
Pr. Nauki 45, Kiev-28, 252650, Ukraine
\ \\
\cite{takis}Concordia University, Montreal, Quebec H3G 1M8, Canada}
\date{\today}
\address{}
\address{\mbox{}}
\address{\parbox{14cm}{\rm \mbox{}\mbox{}\mbox{}
A linear theory of electron transport is developed for a system of
two ideal quantum wires, of length $L$, coupled by tunneling and Coulomb
interaction. The interaction of electrons with acoustical phonons
is included 
and the results are valid in both the ballistic and diffusive regime. In the
{\it ballistic} regime, both tunneling and Coulomb drag lead to a {\it
negative}
transresistance $R_{TR}$, while in the {\it diffusive} regime the
tunneling opposes the drag and leads to a {\it positive} $R_{TR}$.
If $L$ is smaller than the phase-breaking length, the tunneling leads
to interference oscillations of the resistance that are damped
exponentially with $L$.}}
\address{\mbox{}}
\address{\parbox{14cm}{\rm PACS 73.40.Gk, 73.23.-b, 73.23.Ad}}
\maketitle



In submicrometer-long quantum wires and at low temperatures the
electron transport is mainly ballistic$^{1}$ and the wire conductance
reaches its fundamental value of $G_0=e^2/\pi \hbar$. In contrast, for
sufficiently long wires this transport is limited by scattering
processes. The tunneling of electrons between parallel quantum wires
and/or interlayer electron-electron (e-e) interaction are
essential as demonstrated, e.g., by experimental and
theoretical works on electron transport$^{2-13}$ along
the layers in coupled double-wire systems. These works
are mostly devoted to studying interlayer tunneling in the
purely ballistic regime$^{2-10}$ or momentum transfer between
the wires (Coulomb drag)$^{11-13}$. The drag effect has been
studied both in the diffusive$^{11}$ and ballistic$^{12}$
transport regimes as well as 
when the electrons are described by a Luttinger liquid$^{13}$.

Despite this progress, the description of electron transport in
coupled quantum wires is substantially insufficient. Even within
the concept that the electrons are described by a normal Fermi liquid,
two important questions arise. The first one is how to describe the
transport properties when both tunneling and the interactions of
electrons with each other, and with impurities or phonons, are
essential. The second question is how to bridge the gap between
the ballistic and diffusive  regimes in such a description. 
In this Letter we present a linear-response theory of electron
transport in coupled quantum wires that gives a reasonable answer to
both questions and the new results mentioned in the abstract.

{\it From kinetic theory to local description}.
We consider two parallel, tunnel-coupled 1D layers of degenerate
electrons adiabatically contacted to equilibrium reservoirs (leads)
as shown and labelled in Fig. 1. We start from the quantum kinetic
equation $\partial \hat{\rho}/ \partial t + (i/\hbar) [\hat{H}_0 +
\hat{H}_C + \hat{H}_{e-ph}, \hat{\rho}] =0$ for the density matrix
$\hat{\rho}$ and assume that electrons interact with the Coulomb field
($\hat{H}_C$) and with acoustical phonons($\hat{H}_{e-ph}$).
Elastic scattering is neglected, i.e., we assume ideal wires.
We employ a tight-binding description with basis that of the
isolated left ($l$) and right ($r$) layer states in which the potential
energy matrix is $\hat{h}=(\Delta/2) \hat{\sigma}_z +
T \hat{\sigma}_x$ when only the lowest subband is occupied in
either layer. Here $\hat{\sigma}_i$ are the Pauli matrices,
$\Delta$ is the level splitting energy, and $T$  the tunneling
matrix element.

The kinetic equation can be written$^{14}$ as one for the Keldysh's
Green's function $\hat{G}^{-+}$. Below we consider the case when
the characteristic spatial scale of the electronic distribution
is large in comparison with the Fermi wavelength $\pi \hbar/p_F$
and use the Keldysh's matrix Green's function in the Wigner
representation $\hat{G}^{-+}_{{\textstyle \varepsilon, t}}(p,x)$,
where $p$ and $\varepsilon$ are the momentum and energy. In the
steady-state regime the time-averaged Green's functions are
linearized in the manner $\hat{G}^{\alpha \beta}_{{\textstyle
\varepsilon}}(p,x)= \hat{G}^{(0) \alpha \beta}_{{\textstyle
\varepsilon}}(p) + \delta \hat{G}^{\alpha \beta}_{{\textstyle
\varepsilon}}(p,x)$, where $\alpha$ and $\beta$ stand for $+$ or $-$.
The unperturbed part $\hat{G}^{(0) \alpha \beta}_{{\textstyle
\varepsilon}}(p)$ is expressed through the retarded and advanced
matrix Green's functions and the equilibrium Fermi-Dirac function
$f(\varepsilon)=1/[1+e^{(\varepsilon-\mu)/k_BT_e}]$ in the usual
way$^{14}$. The linearized kinetic equation reads

	\begin{eqnarray}
	\nonumber
	\frac{\hbar}{2} \{ \hat{v}_p&,& \frac{\partial}{\partial x}
	\delta \hat{G}^{-+}_{{\textstyle \varepsilon}}(p,x)\} 
	+ i [\hat{h}, \delta \hat{G}^{-+}_{{\textstyle
	\varepsilon}}(p,x) ]  + i [\hat{\varphi}, \hat{G}^{(0)-+}_{{\textstyle
	\varepsilon}}(p) ]\\*
	&&- \frac{\hbar}{2} \{ \frac{\partial}{\partial x}
	\hat{\varphi} , \frac{\partial}{\partial p}
	\hat{G}^{(0)-+}_{{\textstyle \varepsilon}}(p) \}
	 = i \delta \hat{{\cal I}}(\varepsilon, p,x).
	\end{eqnarray}
Here $\{...\}$ denotes anticommutators, $\hat{v}_p=\hat{P}_l
v_{lp}+ \hat{P}_r v_{rp}$ is the diagonal matrix of the group velocities,
and $\hat{P}_l$ and $\hat{P}_r$ are the projection matrices.
We consider only the case of equal group velocities in the layers,
when $v_{lp}=v_{rp}=v_p=p/m$. Further, $\hat{\varphi}$ is the matrix
of the self-consistent electrostatic potential resulting from the
perturbation of the electron density. The collision integral
is given as$^{14}$ $\hat{{\cal I}}= - [ \hat{\Sigma}^{-+} \hat{G}^{++}
+ \hat{\Sigma}^{--} \hat{G}^{-+} + \hat{G}^{-+} \hat{\Sigma}^{++} +
\hat{G}^{--} \hat{\Sigma}^{-+} ]$, where $\hat{\Sigma}^{\alpha \beta}$
is the self-energy due to Coulomb and electron-phonon interactions, and
where the arguments of all functions are $\varepsilon$, $p$, and $x$. This corresponds to a quasiclassical
description of the scattering. Despite the approximations made,
Eq. (1), due to its matrix structure, is not reduced to a classical
Boltzmann equation.

We assume that the Fermi energy is large in comparison
with both $T$ and $\Delta$. We sum up Eq. (1) over the electron
momentum $p$ in the regions of forward ($+$) and backward ($-$)
group velocities and introduce the non-equilibrium part of the
energy distribution function $\hat{g}_{\textstyle \varepsilon}(x)$

    \begin{equation}
    \hat{g}^{\pm}_{\textstyle \varepsilon}(x)=(1/2 \pi i)
    \int_{\pm} d p
    |v_p| \delta \hat{G}^{-+}_{{\textstyle \varepsilon}}(p,x).
    \end{equation}
Since $\delta \hat{G}^{-+}$ is essentially nonzero only in narrow
intervals of energy and momentum near the chemical potential $\mu$
and Fermi momentum, we can replace $|v_p|$  by the Fermi velocity
$v_F$ common to both layers. The result is
	\begin{equation}
	\pm v_F\partial  \hat{g}^{\pm}_{\textstyle \varepsilon}(x)/\partial x
	+ (i/\hbar) [\hat{h}, \hat{g}^{\pm}_{\textstyle \varepsilon}(x) ]
	=   \delta \hat{I}_{\pm}(\varepsilon,x)
	\end{equation}
where $\delta \hat{I}_{\pm}(\varepsilon,x)=(1/2 \pi \hbar)\int_{\pm}
|v_p| d p~\delta \hat{{\cal I}}(\varepsilon, p,x)$ depends on both
$\hat{g}^{+}$
and $\hat{g}^{-}$ since it accounts for both forward- and backscattering
processes. Notice that the absence of $\hat{\varphi}$ in Eq. (3) is not a
simplification or an approximation: it results exactly from the 
integration of Eq. (1) over momentum.

The matrix kinetic equation (3) is equivalent to eight scalar equations for
the four components of $\hat{g}^{+}$ and those (four) of $\hat{g}^{-}$. The
boundary conditions for them are determined, in the Landauer-B{\"u}ttiker-Imry
approach, by the distribution functions of the leads described by four
chemical potentials $\mu_{1l}$, $\mu_{1r}$, $\mu_{2l}$, and $\mu_{2r}$,
so that $\hat{g}^{+}_{\textstyle \varepsilon}(0)=
- (\partial f(\varepsilon)/\partial
\varepsilon) [\hat{P}_l \delta \mu_{1l} + \hat{P}_r
\delta \mu_{1r} ],$ and $\hat{g}^{-}_{\textstyle \varepsilon}(L)=
- (\partial f(\varepsilon)/\partial
\varepsilon) [\hat{P}_l \delta \mu_{2l} + \hat{P}_r \delta \mu_{2r} ].$
The forward- and backward-propagating states are "connected" to
leads "1" and "2", respectively. The nondiagonal components vanish
at the contacts since the tunneling is absent beyond the region
$x=[0,L]$.

We now introduce the local chemical potential matrix
$\delta \hat{\mu}^{\pm}(x)= \hat{P}_l \delta \mu_l^{\pm}(x)+
\hat{P}_r \delta \mu_r^{\pm}(x) + \hat{\sigma}_x \delta \mu_x^{\pm}(x)
+\hat{\sigma}_y \delta \mu_y^{\pm}(x)$ according to
$\delta \hat{\mu}^{\pm}(x)=\int d
\varepsilon ~\hat{g}^{\pm}_{\textstyle \varepsilon}(x)$. Assuming that
backscattering occurs much less frequently than forward-scattering,
which is the case when the e-e scattering dominates over the electron-phonon
one, and that tunneling also occurs much less frequently
than forward-scattering, we can write $[\hat{g}_{\textstyle
\varepsilon}^{\pm}(x)]_{ll,rr}= - (\partial f(\varepsilon)/\partial
\varepsilon) \delta \mu_{l,r}^{\pm}(x)$. Below we omit the symbol
"$\delta$" in $\delta \hat{\mu}^{\pm}(x)$ and in the potentials
of the leads since all potentials are measured from the same equilibrium
value $\mu$. Then integrating Eq. (3) over the energy we obtain eight
coupled, first-order differential equations:

	\begin{eqnarray}
	\nonumber
	\pm d \mu_l^{\pm}/dx + (\mu_l^{\pm}-\mu_l^{\mp})(1/l_P+1/l_D)\\ 
	- (\mu_r^{\pm} -\mu_r^{\mp})/l_D - 2t_F  \mu_y^{\pm} =0,\\
	\nonumber
	\ \\
	\nonumber
	\pm d \mu_r^{\pm}/dx + (\mu_r^{\pm} -\mu_r^{\mp})(1/l_P+1/l_D) \\
	- (\mu_l^{\pm}
	-\mu_l^{\mp})/l_D + 2t_F  \mu_y^{\pm} =0,\\ 
	\nonumber
	\ \\
	\pm d \mu_x^{\pm}/dx + \delta_F \mu_y^{\pm} + \mu_x^{\pm}/l_C=0, \\
	\pm d \mu_y^{\pm}/dx - \delta_F \mu_x^{\pm} + \mu_y^{\pm}/l_C +
	t_F  (\mu_l^{\pm} -\mu_r^{\pm}) =0.
	\end{eqnarray}
Here $t_F= T/\hbar v_F, \delta_F=\Delta/\hbar v_F$. The evaluation
of $\delta \hat{I}$ was carried out assuming a weak tunnel coupling, when $T$
is small in comparison to the imaginary part of the self-energies.
The characteristic lengths $l_P$, $l_D$ and $l_C$, resulting from the
collision integral $\delta \hat{I}_{\pm}(\varepsilon,x)$,
are expressed, respectively, through the phonon-assisted 1D transport
time$^{15}$ $\tau_P$, the 1D Coulomb drag time$^{11}$ $\tau_D$, and the
phase-breaking time $\tau_C$ (which describes the suppression of tunnel
coherence) as $l_P = 2 v_F \tau_P$, $l_D = 2 v_F \tau_D$, and $l_C =
v_F \tau_C$. All these times are microscopically justified.  
For $\tau_C$, if we account  only for the lowest-order
Coulomb contributions, we obtain the new result

	\begin{equation}
	1/\tau_{C} \simeq [e^4 \ln^2 (w/a) \Delta/2 \pi \hbar^3
	\epsilon^2 v_F^2] cotanh  (\Delta/4 k_BT_e)
	\end{equation}
where $\epsilon$ is the dielectric constant, $w$ is the distance
between the wires and $a$  the wire width. Estimating $\tau_{C}$ from
Eq. (8), we find  $l_C \ll l_P$ and $l_C \ll l_D$, because of the
weakness of the electron-phonon coupling and of the e-e backscattering,
respectively.

Equations (4)-(7) with the boundary conditions
$\mu_l^+(0)=\mu_{1l}$, $\mu_l^-(L)=\mu_{2l}$, $\mu_r^+(0)=\mu_{1r}$,
$\mu_r^-(L)=\mu_{2r}$ and $\mu_{x,y}^+(0)=\mu_{x,y}^-(L)=0$
give  a complete description
of  electron transport in double quantum wire systems in a wide range of
regimes starting from the purely ballistic  regime, for $L \ll l_C$, to
the diffusive regime, for $L \gg l_P, l_D$. The local currents flowing
in the layers $j=l,r$ are expressed by the Landauer-like formula
\begin{equation}
J_{j}(x)= G_0 [ \mu_j^+(x) - \mu_j^-(x) ]/e.
\end{equation}
Below, to characterize the effects of drag and tunneling, we
calculate the transresistance $R_{TR}=[\mu_{1l}-\mu_{2l}]/eJ_r$,
where $J_r$ is the current injected in the wire $r$ when no current
is allowed to flow into wire $l$. This is the typical setup
for the drag experiments. In this case $J_r(0)=J_r(L)=J_r$ and
$J_l(0)=J_l(L)=0$. We also calculate the direct resistance
$R=[\mu_{1r}-\mu_{2r}]/eJ_r$.

{\it Results}. The solution of Eqs. (4)-(7) is obtained for all potentials
in the form of $A + Bx + \sum_{\pm}C_i^{\pm} e^{\pm \lambda_i x}$,
where $\lambda_i^2$ are the roots of a cubic equation. Explicitly,
for $l_C \ll l_P,l_D$ and weak tunnel coupling $t_F \ll l_C^{-1}$, the
three roots are $\lambda_1=\lambda \simeq 2 (1/l_T^2+1/l_T l_1)^{1/2}$
and $\lambda_{2,3}=\lambda_{\pm} \simeq 1/l_C \pm i \delta_F$,
where $1/l_1=1/l_P+2/ l_D$ and $l_T=v_F \tau_T$. The
tunneling time $\tau_T$ is given by $1/\tau_T =( 1/\tau_C) 2T^2/(\Delta^2+
\hbar^2/\tau_C^2 )$. The terms with $\lambda$ describe long-scale
variations of the chemical potentials, while those with $\lambda_{\pm}$
correspond to short-scale variations. Two cases can be
considered.

{\it Long wires, $L \gg l_C$ }.
The solutions containing $\lambda_{\pm}$ 
are not essential in the calculation of the currents. Considering
only the solutions with $\lambda$, we find

\begin{eqnarray}
R = (\pi \hbar/ 2 e^2) [ 2 +  L/l_P + \sqrt{1+l_T/l_1} \tanh(\lambda L/2)],\\
R_{TR}= (\pi \hbar/ 2 e^2) [L/l_P -  \sqrt{1+l_T/l_1}\tanh(\lambda L/2)].
\end{eqnarray}
In the {\it ballistic} regime ($L \ll l_P, l_T$) we obtain the usual resistance
$R \simeq  \pi \hbar/e^2=G_0^{-1}$, while the
transresistance is given by $R_{TR}=-(\pi \hbar /e^2)L [1/l_D + 1/2 l_T]$.
As seen, $R_{TR}$ is small, always {\it negative}, and proportional to $L$.
If we neglect tunneling, the resulting $R_{TR}$ describes the Coulomb drag in 
the ballistic regime$^{12}$. In the {\it diffusive} regime, i.e.,
when $L$ increases and the  relation $L \gg l_P$ holds, we obtain
$R \simeq (\pi \hbar /e^2) L [1/l_P+1/l_D]$ for $\lambda L/2 \ll 1$. 
This resistance, if one omits the drag contribution, is expressed in terms of 
the  Drude conductivity $\sigma= L/R=e^2 l_P/\pi \hbar = e^2 n \tau_P/m$,
where $n$ is the 1D electron density in layer $r$. 
Then $R_{TR}=-(\pi \hbar/e^2)(L/l_D) [1 - (L/L_0)^2]$ where
$L_0=(6 l_P^2 l_T/l_D)^{1/2}$. Introducing the drag transresistivity
 $\pi \hbar /e^2 l_D$ and the 1D-1D tunneling conductance $e^2
\rho_{1D}/\tau_T = 2 e^2/\pi \hbar l_T$, where $\rho_{1D}$ is the 1D
density of states at the Fermi level, we formally obtain the result of
Ref. 16  where a competition of drag and tunneling  was investigated,
for double quantum well systems,  only in the {\it diffusive} regime. 
As the factor $[1 - (L/L_0)^2]$ shows,
in this  regime the tunneling opposes the drag and the 
transresistance increases, changing its sign from negative to positive. If $\lambda L/2 \sim 1$, $R_{TR}$ is large and always comparable to $R$, because a 
considerable fraction of the current penetrates in the $l$ layer due to 
tunneling. In Fig. 2 we show the length dependence of $R_{TR}$
given by Eq. (11) for different relative contributions of the Coulomb
drag and tunneling. $R_{TR}$  is {\it negative} for small $L$ but it always
changes sign and becomes {\it positive} as $L$ increases. 
This occurs at smaller $L$ when the tunneling is stronger (larger $l_P/l_T$)
or the drag is weaker (smaller $l_P/l_D$).

A peculiar transport regime, without backscattering, can be realized in
tunnel-coupled magnetic edge states$^{17,18}$. Assuming $1/l_P=1/l_D=0$
in Eqs. (10) and (11), we obtain the result of Ref. 18 in the form
$R = (\pi \hbar /e^2) \left[ 1 + (1/2)\tanh (L/l_T) \right]$ and
$R_{TR} = -(\pi \hbar /2 e^2) \tanh (L/l_T)$.


{\it Short wires $L \sim l_C \ll l_P, l_D$}. Electrons pass
through the wires almost without  backscattering,
$R \simeq \pi \hbar /e^2$, and $R_{TR}$ is small. However,
an electron tunneling between the layers does not lose its
phase memory completely; as a result tunnel coherence effects can
take place 
and give additional contributions to $R$
and $R_{TR}$:

	\begin{eqnarray}
	R = {\pi \hbar\over e^2} [ 1 + {L\over l_P}  +
	{L\over l_D} + {L\over 2l_T}
	 -{l_C\over 2 l_T} \Phi(L)],\\
	\nonumber
	\ \\
	R_{TR} = {\pi \hbar \over e^2}[ -{L\over l_D} - {L\over 2l_T }+
	{l_C \over 2 l_T} \Phi(L)],
	\end{eqnarray}
where
	\begin{eqnarray}
	\nonumber
	\Phi(L)&=&(l_C^{-2}+\delta_F^2)^{-1} [ 2 (\delta_F /l_C)
	e^{-L/l_C} \sin (\delta_F L)\\*
	&& + (l_C^{-2}-\delta_F^2)
	 (1-e^{-L/l_C} \cos (\delta_F L) ) ].
	\end{eqnarray}

The term proportional to $\Phi(L)$ describes oscillations of the resistance
damped due to the factor $\exp (-L/l_C)$. Its relative contribution to
$R_{TR}$ is not small at $L \sim l_C$. The periodic behavior can be
described as a result of tunneling-assisted interference of electron
waves of the left and right layers on the coupling length $L$. Due to
a finite $\Delta$, these waves have slightly different phase velocities.
Figure 3 shows the dependence of $\Phi(L)$ as a function of $\delta_F L$;
we used Eq. (8) at zero temperature so that $1/l_C= \kappa |\delta_F|$,
where $\kappa$ is a dimensionless parameter.

The level splitting $\Delta$  can be changed
not only by a transverse voltage, but also by a  magnetic field$^9$
$H$ perpendicular to the plane of the  quantum wires. For sufficiently
weak $H$ the results presented so far still hold with the phase
$\delta_F L$ in Eq. (14)  changed to $\delta_F L+ 2 \pi \phi/\phi_0$,
where $\phi_0=h/e$ is the magnetic flux quantum and $\phi = H w L$  the
flux enclosed by the area between the wires. Though the double-wire system
does not form a closed current loop, this should lead
to Aharonov-Bohm-type oscillations in $R$ and $R_{TR}$.

In very short wires, with $L \ll l_C, \delta_F^{-1}$,
we have $R_{TR} = (\pi \hbar/e^2)[ -L/l_D - L^2 t_F^2/2]$: in
contrast with the drag contribution, 
that of tunneling  shows a $L^2$ dependence.

Finally, we apply Eq. (3) to the {\it purely ballistic} regime in which
we can neglect the collision integral and need not make
the assumption of weak tunnel coupling. Then we obtain Eqs.
(4)-(7) without the terms containing $l_P$, $l_D$ and $l_C$; they
describe oscillations of the electronic wave packets between
the layers due to coherent tunneling. Solving them, we obtain
	\begin{eqnarray}
	R = (\pi \hbar/e^2) [ 1  
	-r\sin^2 \psi] [ 1 - 2r 
	\sin^2 \psi ]^{-1}, \\
	R_{TR} = -(\pi \hbar/e^2) r 
	\sin^2 \psi [ 1 -2r 
	\sin^2 \psi ]^{-1},
        \end{eqnarray}
where $r=2T^2/\Delta_T^2$, $\Delta_T=(\Delta^2 +4 T^2)^{1/2}$ and
$\psi=\Delta_T L/2 \hbar v_F$.
The periodicity of the oscillations becomes the same as in Eqs. (12) and
(13) if one replaces $\Delta$ by $\Delta_T$. However, since the tunnel
coupling is not weak, the oscillations described by Eqs. (15) and (16)
have large amplitudes. In particular, when $\Delta$ is small, both $R$
and $R_{TR}$ show giant oscillations  with  amplitude large in
comparison with $1/G_0$. The experimentally
observed$^6$ resistance oscillations,
resulting from tunnel coupling over a {\it finite-length} region, 
in  ballistic quantum wires were of small ($\sim$ 0.5 K$\Omega$) amplitude. This is
not surprising because there are many factors, e.g., long-scale
inhomogeneities, inelastic and elastic scattering, which compete
against the tunnel coherence. Previous theoretical studies
of the purely ballistic transport regime$^{2-5,7-9}$ were
based on a quantum-mechanical calculation of the electronic wave
transmission, while we recover the essential results from a
quantum-kinetic analysis.

In conclusion, we developed a linear-response, steady-state theory of
electron transport in parallel one-dimensional layers coupled by
tunneling and Coulomb interaction and contacted to
quasi-equilibrium reservoirs. The quantum-kinetic description of the problem 
leads to linear differential equations describing the distribution
of local chemical potentials of the forward- and backward-moving electrons. 
The results for the direct resistance $R$ and the transresistance $R_{TR}$ 
are valid in the whole range from the pure ballistic to the diffusive regime. 
$R_{TR}$ always depends nonmonotonically on the wire length $L$:
in the {\it ballistic} regime tunneling and  drag lead to a
{\it negative} transresistance $R_{TR}$, while in the {\it diffusive} regime
the tunneling opposes the drag and leads to a {\it positive} $R_{TR}$.
When $L$ is smaller than  the phase-breaking length, the tunneling leads to
interference oscillations of the resistance damped exponentially with $L$.

This work was supported by the Canadian NSERC Grant No. OGP0121756.

\end{document}